\documentclass[aps,pra,preprint,superscriptaddress,amsmath,amssymb,showpacs]{revtex4-1}
\usepackage{bm}

\begin{document}

\title{Ultrarelativistic quasiclassical wave functions in strong laser and atomic fields}

\author{A. Di Piazza}
\email{dipiazza@mpi-hd.mpg.de}
\affiliation{Budker Institute of Nuclear Physics of SB RAS, 630090 Novosibirsk, Russia}
\affiliation{Max-Planck-Institut f\"ur Kernphysik, Saupfercheckweg 1, 69117 Heidelberg, Germany}

\author{A. I. Milstein}
\email{milstein@inp.nsk.su}
\affiliation{Budker Institute of Nuclear Physics of SB RAS, 630090 Novosibirsk, Russia}

\date{\today}

\begin{abstract}
The  problem of an ultrarelativistic charge in the presence of an atomic and a plane-wave field is investigated in the quasiclassical regime by including exactly the effects of both background fields. Starting from the quasiclassical Green's function obtained in [Phys. Lett. B \textbf{717}, 224 (2012)], the corresponding in- and out-wave functions are derived in the experimentally relevant case of the particle initially counterpropagating with respect to the plane wave. The knowledge of these electron wave functions opens the possibility of investigating a variety of problems in strong-field QED, where both the atomic field and the laser field are strong enough to be taken into account exactly from the beginning in the calculations.
\end{abstract}

\maketitle

\section{Introduction}

Quantum Electrodynamics (QED) is certainly the most extensively tested physical theory and the agreement between experimental and theoretical results is outstanding \cite{Hanneke_2008}. The experimental validation of the predictions of QED in the presence of external strong electromagnetic fields is, however, less thorough. In the realm of QED, an electric (magnetic) field is denoted as ``strong'' if it is of the order of the so-called critical electric (magnetic) field of QED $E_c=m^2/|e|=1.3\times 10^{16}\;\text{V/cm}$ ($B_c=m^2/|e|=4.4\times 10^{13}\;\text{G}$), where $e$ and $m$ are the electron charge and mass, respectively (units with $\hbar=c=1$ are employed throughout the paper) \cite{Sauter_1931,Heisenberg_1936,Schwinger_1951,Landau_b_4_1982}. Now, the lack of a thorough experimental investigation of strong-field QED is due, on the experimental side, to the difficulty of producing such strong electromagnetic fields and, on the theoretical side, to the complexity of the underlying methods, where the background electromagnetic field has to be taken into account exactly from the beginning.

An important exception to the mentioned experimental difficulty is represented by the electromagnetic field of highly-charged ions, i.e., of ions with charge-number $Z$ such that $Z\alpha\sim 1$, with $\alpha=e^2\approx 1/137$ being the fine-structure constant. The electric field of such ions, indeed, is of the order of $E_c$ at a distance from the charge of the order of the typical QED ``Compton'' length $\lambda_C=1/m\approx 3.9\times 10^{-11}\;\text{cm}$. This explains why QED processes occurring in the presence of highly-charged ions have been investigated theoretically already since a long time \cite{Bethe_1934} (see also the reviews \cite{Milstein_1994,Lee_2003,Baur_2007,Baltz_2008}). Correspondingly, a large number of experiments have been performed, confirming the prediction of QED and, as prominent examples, we quote the experiments on the $(g-2)$-factor of bound electrons \cite{Sturm_2011} and on vacuum-polarization effects primed by highly-charged ions as Delbr\"{u}ck scattering \cite{Delbexp} and photon splitting \cite{splitexp}.

Presently available laser systems are able to deliver peak electromagnetic fields, which are still about four orders of magnitude below $E_c$ and $B_c$ \cite{Yanovsky_2008}. However, the ratio between the laser electric-field amplitude $E$ and $E_c$ in the laboratory frame is not a Lorentz invariant parameter and, as it is, it cannot enter physical observables. Indeed, if a QED process in the presence of a laser field involves an incoming counterpropagating particle with energy $\epsilon\gg m$, the physically relevant parameter for estimating nonlinear QED effects is the quantity $\chi=(\epsilon/m)(E/E_c)$ \cite{Ritus_1985,Baier_b_1998,Di_Piazza_2012}. In the case of an incoming electron or positron, the parameter $\chi$ represents (half of) the electric-field amplitude of the laser in the initial rest frame of the particle in units of $E_c$. Therefore, the strong-field QED regime $\chi\sim 1$ can be effectively entered with present laser technology by employing, for example, ultra-relativistic electron beams with a few GeV energy. Such high-energy electron beams are already available both in conventional accelerators \cite{PDG_2012} and in laser-plasma accelerators \cite{Wang_2013}. The influence of a laser field on a QED process is also characterized by the so-called classical intensity parameter $\xi=|e|E/m\omega_0$, with $\omega_0$ being the typical angular frequency of the laser field \cite{Ritus_1985,Baier_b_1998,Di_Piazza_2012}. Presently available optical laser systems allow for values of $\xi$, which largely exceed unity, implying that relativistic and multiphoton effects stemming from the laser field have to be taken into account exactly in the study of corresponding QED processes \cite{Ritus_1985,Baier_b_1998,Di_Piazza_2012}.

The high intensity of present-day lasers also opens the possibility of investigating the influence of laser light on such QED processes in an atomic field as electron-positron ($e^+e^-$) photoproduction (Bethe-Heitler process) and bremsstrahlung. Various publications have been devoted to the study of the influence of a laser field on QED processes occurring in an atomic field including exactly the laser field but only in the leading approximation in the parameter $Z\alpha$ (Born approximation). For example, the Bethe-Heitler process was considered in \cite{Loetstedt_2008} and in \cite{Di_Piazza_2009}, whereas bremsstrahlung was investigated in \cite{Loetstedt_2007} and Delbr\"{u}ck scattering in \cite{Di_Piazza_2008}. Since the mentioned results have been obtained in the leading approximation in $Z\alpha$, only the electron states exact in the laser field (Volkov states) have been employed \cite{Landau_b_4_1982}. In \cite{DPM2012}, we have determined the quasiclassical Green's function of the Klein-Gordon and of the Dirac equation in the ultrarelativistic regime, by including exactly both the atomic field (via the parameter $Z\alpha$) and the laser field (via the parameters $\xi$ and $\chi$). Although the Klein-Gordon and the Dirac equations cannot be solved exactly in the combined atomic and laser field, this has been achieved by means of the operator technique based on the quasiclassical Green's function in the leading order with respect to the parameter $m/\epsilon\ll 1$. This technique has been already applied for the case of an arbitrary spherically symmetric field \cite{LM95A}, and for a localized field which generally possesses no spherical symmetry \cite{LMS00}. Different applications of the operator technique to the investigation of high-energy QED processes in background fields, where the Dirac equation cannot be solved analytically, can also be found in \cite{Baier_b_1998}. In this respect, we mention here that the operator technique was first developed to investigate QED processes in the presence of a background field where the corresponding Klein-Gordon and Dirac equations can be solved exactly, as an alternative method which does not require the use of the explicit form of the electron's wave functions in the field (see \cite{Schwinger_1951,Baier_1974,Baier_1975} for the case of a constant homogeneous electromagnetic field, \cite{Baier_1975_b,Baier_1975_c} for the case of a plane electromagnetic wave, and \cite{Milstein_1982} for the case of a Coulomb field).

In \cite{DPM2012}, we have also applied the quasiclassical Green's function to calculate the correction induced by the presence of the laser field to the total Bethe-Heitler pair photoproduction cross section in the ultrarelativistic regime in the leading order with respect to the parameter $m/\epsilon\ll 1$. However, the availability of the Green's function alone does not allow for the calculation of differential cross sections, which give information, for example, on the angular and/or on the energy distribution of final particles. In order to achieve this goal, the knowledge of the particle's wave function in the corresponding combined atomic and plane-wave field is required and, in the present paper, we will solve this problem. In \cite{LMS00}, it was shown how it is possible to apply a limiting procedure, to obtain the particle wave functions from the corresponding Green's function in the case of a background atomic field. Here, we extend this technique to the case where also a plane-wave field is present, which, as in \cite{DPM2012}, counterpropagates with respect to the particle. Analogously to as in \cite{DPM2012}, the present results are valid for particle energies $\epsilon\gg m$ but such that $\omega_0\epsilon\ll m^2$. We also assume that  $Z\alpha < 1$,  $\chi\lesssim 1$, and  $\xi\gg 1$. However, in order to make sure that the particle is only barely deviated from the initial direction by the laser field, we also require that $m\xi\ll \epsilon$, i.e., that the typical transverse momentum of the particle in the laser field is much smaller than its longitudinal one, which in turn is of the order of the particle energy.

The paper is organized as follows. In Sec. II a thorough derivation of the Green's functions of the Klein-Gordon and Dirac equations in the presence of an atomic and a plane-wave field is reported. Although the final results of the Green's functions can already be found in \cite{DPM2012}, details of the derivation are presented, which also help the reader in further understanding the region of validity of the Green's functions themselves. In Sec. III the corresponding detailed derivation of the pair photoproduction cross section including the effects of the laser field, already reported in \cite{DPM2012}, is also presented. In Sec. IV the obtained Green's functions are employed to determine the corresponding scalar and spinor quasiclassical wave functions of an ultrarelativistic charge counterpropagating with respect to the laser field. Finally, in Sec. V the main conclusions of the paper are presented.

\section{Detailed derivation of the Green's function}
As we have mentioned in the introduction, in this section we will derive in more detail the Green's functions already presented in \cite{DPM2012}.

We consider an ultrarelativistic particle with mass $m$ and with momentum directed almost along the positive $z$-direction in the presence of a counterpropagating plane wave described by the vector potential $\bm{A}(t+z)$, with $\bm{z}\cdot\bm{A}(t+z)=0$, and of an atomic field described by the scalar potential $V(\bm{r})$, localized around the origin of coordinates. It is convenient to pass from the variables $t$ and $z$ to the light-cone variables $\phi=t-z$ and $T=(t+z)/2$, and to denote as $\bm\rho$ the component of the vector $\bm r$ perpendicular to $\bm z$. In this way
\begin{eqnarray}
&& p^0=i\partial_t=-p_\phi-p_T/2\, ,\quad
p^z=-i\partial_z=-p_\phi+p_T/2\,,\nonumber\\
&&p_\phi=-i\partial_\phi\,,\quad p_T=-i\partial_T\,, \quad \bm p_\perp=-i\partial_{\bm\rho}\,.
\end{eqnarray}
We first consider the case of a scalar particle and determine the Green's function of the Klein-Gordon equation. In the ultrarelativistic regime it is possible to replace $z$ in the argument of the potential $V(\bm{r})$ by $T$, i.e., $V(\bm{r})=V(\bm{\rho},T)$, whereas by definition it is $\bm{A}(t+z)=\bm{A}(2T)$. Since the total background field is independent of the variable $\phi$, the Green's function of the Klein-Gordon equation can be written in the form
\begin{equation}\label{KGV}
D^{(0)}(\phi,\,T,\,\bm\rho|\, \phi',\,T',\,\bm\rho')=\int\frac{d\epsilon}{2\pi}\mbox{e}^{-i\epsilon(\phi-\phi')}
D^{(0)}(T,\,\bm\rho|\, T',\,\bm\rho';\,\epsilon),
\end{equation}
where
\begin{equation}
\begin{split}
D^{(0)}(T,\,\bm\rho|\, T',\,\bm\rho';\,\epsilon)&=
\frac{1}{{\cal P}^2-m^2+i0}\delta(T-T')\delta(\bm\rho-\bm\rho')\\
&=-i\int_0^\infty ds\,\exp\left[is({\cal P}^2-m^2)\right]\delta(T-T')\delta(\bm\rho-\bm\rho'),
\end{split}
\end{equation}
with
\begin{equation}
\begin{split}
{\cal P}^2&=(\epsilon-p_T/2-V(\bm\rho,\,T))^2-(\epsilon+p_T/2)^2-(\bm p_\perp-\bm{\mathcal{A}}(T))^2\\
&=-\left[2\epsilon p_T+2\epsilon V(\bm\rho,\,T)+(\bm p_\perp-\bm{\mathcal{A}}(T))^2-\frac{1}{2}\{p_T,\,V(\bm\rho,\,T)\}-V^2(\bm\rho,\,T)\right].
\end{split}
\end{equation}
Here, we set for simplicity $\bm{\mathcal{A}}(T)=e\bm A(2T)$. Below, in the expression of ${\cal P}^2$ we neglect the term $V^2(\bm\rho,\,T)$ in comparison with $\epsilon V(\bm\rho,\,T)$ and the term $\{p_T,V(\bm{\rho},T)\}=p_T\,V(\bm{\rho},T)+V(\bm{\rho},T)\,p_T$ in comparison with $\epsilon p_T$.

\subsection{The case of vanishing plane-wave field}
In order to demonstrate the operator method of calculation, we start from the simpler case where $\bm {\mathcal A}(T)=\bm{0}$. Let us consider the function
\begin{eqnarray}\label{KGV1}
F(T,\,\bm\rho|\, T',\,\bm\rho')=
\mbox{e}^{-is[2\epsilon V(\bm\rho,\,T)+H_{0}]}\delta(T-T')\delta(\bm\rho-\bm\rho'),
\end{eqnarray}
where $H_{0}=2\epsilon p_T+\bm p_\perp^2$. In terms of this function, the Green's function $D^{(0)}(T,\,\bm\rho|\, T',\,\bm\rho';\,\epsilon)$ reads
\begin{equation}\label{KGVF}
 D^{(0)}(T,\,\bm\rho|\, T',\,\bm\rho';\,\epsilon)=-i\int_0^\infty ds\,\mbox{e}^{-ism^2}F(T,\,\bm\rho|\, T',\,\bm\rho').
\end{equation}
Now, we write the function $F(T,\,\bm\rho|\, T',\,\bm\rho')$ as
\begin{equation}\label{F}
F(T,\,\bm\rho|\, T',\,\bm\rho')=L_0(s)\,\mbox{e}^{-isH_{0}}\delta(T-T')\delta(\bm\rho-\bm\rho'),
\end{equation}
with
\begin{equation}\label{L}
L_0(s)=\mbox{e}^{-is[2\epsilon V(\bm\rho,\,T)+H_{0}]}\mbox{e}^{isH_{0}}.
\end{equation}
By exploiting the relations
\begin{align}
\label{exp_1}
\mbox{e}^{-i\tau p_T} f(T)&=f(T-\tau)\mbox{e}^{-i\tau p_T},\\
\label{exp_1_p}
\mbox{e}^{-i\beta \bm p_\perp^2} g(\bm\rho)&=\int\frac{d\bm{q}}{i\pi}\mbox{e}^{iq^2}g(\bm\rho+2\sqrt{\beta}\bm q)
\end{align}
for $\beta>0$ and valid for arbitrary functions $f(T)$ and $g(\bm{\rho})$, with $\bm{q}$ being a two-dimensional vector
perpendicular to the $z$-direction (see the Appendix for a derivation of the second, less trivial, relation), it is easy to show that
\begin{equation}\label{H1}
\mbox{e}^{-isH_{0}}\delta(T-T')\delta(\bm\rho-\bm\rho')=-\frac{i}{4\pi s}\delta(T-T'-2\epsilon s)\exp\left[i\frac{(\bm\rho-\bm\rho')^2}{4s}\right].
\end{equation}
In this way, the operator $L_0(s)$ is found to fulfill the equation
\begin{equation}
\frac{d}{ds}L_0(s)=L_0(s)\mbox{e}^{-isH_{0}}[-2i\epsilon V(\bm\rho,\,T)]\mbox{e}^{isH_{0}}
=-2i\epsilon L_0(s)V(\bm\rho-2 s\bm p_\perp,\,T-2\epsilon s),
\end{equation}
where we have also employed the general relation
\begin{equation}
\label{exp_p2}
\mbox{e}^{-i\beta \bm p_\perp^2} g(\bm\rho)=g(\bm\rho-2\beta\bm{p}_{\perp})\mbox{e}^{-i\beta \bm p_\perp^2}.
\end{equation}
In the leading quasiclassical approximation, we can neglect the non-zero commutator of the operators
$V(\bm\rho-2 s\bm p_\perp,\,T-2\epsilon s)$ at different values of $s$. As a result, it is
\begin{eqnarray}\label{L1}
&&L_0(s)=\exp\left[-2is\epsilon\int_0^1 V(\bm\rho-2 sx\bm p_\perp,\,T-2\epsilon sx)\,dx\right]\,.
\end{eqnarray}
By substituting Eqs. (\ref{L1}) and (\ref{H1}) in Eq. (\ref{F}), we obtain
\begin{eqnarray}\label{F1}
&& F(T,\,\bm\rho|\, T',\,\bm\rho')=-\frac{i}{4\pi s}\delta(T-T'-2\epsilon s)\exp\left[i\frac{(\bm\rho-\bm\rho')^2}{4s}\right]\nonumber\\
&&\times\exp\left[-2is\epsilon\int_0^1 V\Big(\bm\rho-x(\bm\rho-\bm\rho')-2 sx\bm p_\perp,\,T-x(T-T')\Big)\,dx\right]\times 1\,,
\end{eqnarray}
where, for the sake of clarity, we have indicated as $1$ the constant function of unit value. Now, we focus our attention on the quantity
\begin{equation}
\exp\left[-2is\epsilon\int_0^1 V\Big(\bm\rho-x(\bm\rho-\bm\rho')-2 sx\bm p_\perp,\,T-x(T-T')\Big)\,dx\right]\times 1.
\end{equation}
By means of Eq. (\ref{exp_p2}), we obtain
\begin{equation}
\label{exp_2}
\begin{split}
&\int_0^1 V\Big(\bm\rho-x(\bm\rho-\bm\rho')-2 sx\bm p_\perp,\,T-x(T-T')\Big)\,dx\\
&\quad=\int_0^1 \mbox{e}^{-isx \bm p_\perp^2/(1-x)}V\Big(\bm\rho-x(\bm\rho-\bm\rho'),\,T-x(T-T')\Big)\mbox{e}^{isx \bm p_\perp^2/(1-x)}\,dx\,.
\end{split}
\end{equation}
Since in this expression all the derivatives corresponding to the operator $\bm{p}_{\perp}$ act only on the atomic potential, the exponential operators $\exp[\pm isx \bm p_\perp^2/(1-x)]$ give a non-negligible contribution only in the vicinity of the origin, where the atomic potential is localized. In addition,  the formation region of QED processes involving an ultrarelativistic particle (antiparticle) propagating from $(T',\bm{\rho}')$ to $(T,\bm{\rho})$ is characterized by $T,-T'>0$ and $T,|T'|\sim (\epsilon/m)\lambda_C\gg \lambda_C$, or $-T,T'>0$ and $|T|,T'\sim (\epsilon/m)\lambda_C\gg \lambda_C$. Indeed, one can see from the argument $T-x(T-T')$ in the atomic potential in Eq. (\ref{exp_2}) that if $T$ and $T'$ have the same sign then, since $0\le x\le 1$ the particle (antiparticle) never crosses the region where the localized atomic field is strong. These considerations allow one to take into account the operators $\exp[\pm isx \bm p_\perp^2/(1-x)]$ in Eq. (\ref{exp_2}) only for $x$ close to the value $x_0$ such that $T-x_0(T-T')=0$ for $0<x_0<1$ and to omit them otherwise. Thus, going back to Eq. (\ref{F1}), we can rewrite it as
\begin{eqnarray}\label{F2}
&& F(T,\,\bm\rho|\, T',\,\bm\rho')=-\frac{i}{4\pi s}\delta(T-T'-2\epsilon s)\exp\left[i\frac{(\bm\rho-\bm\rho')^2}{4s}\right]
\exp\left[-is\frac{x_0}{1-x_0}\bm p_\perp^2\right]\nonumber\\
&&\times\exp\left[-2is\epsilon\int_0^1 V\Big(\bm\rho-x(\bm\rho-\bm\rho'),\,T-x(T-T')\Big)\,dx\right]\,,
\end{eqnarray}
where now the symbol $\times 1$ has been omitted, as no ambiguity can arise. Finally, by exploiting again the relation in Eq. (\ref{exp_1_p}), we arrive at the final expression of the function $F(T,\,\bm\rho|\, T',\,\bm\rho')$:
\begin{equation}\label{Ffinal}
\begin{split}
&F(T,\,\bm\rho|\, T',\,\bm\rho')=-\frac{1}{4\pi^2 s}\delta(T-T'-2\epsilon s)\exp\left[i\frac{(\bm\rho-\bm\rho')^2}{4s}\right]\\
&\quad\times\int d\bm{q}\,\exp\left[iq^2-2is\epsilon\int_0^1 V\Big(\bm\rho-x(\bm\rho-\bm\rho')+2\sqrt{x_0(1-x_0)s}\,\bm q\,,\,T-x(T-T')\Big)\,dx\right].
\end{split}
\end{equation}
In this expression, there are no momentum operators. By substituting the function $F(T,\,\bm\rho|\, T',\,\bm\rho')$ into Eq. (\ref{KGVF}) and by taking the integral over $s$, we finally obtain the following expression of the Green's function of the Klein-Gordon equation:
\begin{equation}\label{D0final}
\begin{split}
&D^{(0)}(T,\,\bm\rho|\, T',\,\bm\rho';\,\epsilon)=\frac{i\theta(s_0)}{4\pi^2|T-T'|}\exp\left[i\frac{(\bm\rho-\bm\rho')^2}{4s_0}-im^2s_0\right]\\
&\times\int d\bm{q}\,\exp\left[iq^2-2is_0\epsilon\int_0^1 V\Big(\bm \rho-x(\bm \rho-\bm \rho')+2\sqrt{\beta}\,\bm q,T-x(T-T')\Big)\,dx\right],
\end{split}
\end{equation}
where $s_0=(T-T')/2\epsilon$, $x_0=T/(T-T')$, and $\beta=x_0(1-x_0)s_0$. We remind that the term with $\bm q$ in the atomic potential should be omitted if $x_0<0$ or $x_0>1$. In this respect, we observe here that the origin of the appearance of this term lies in the non-commutativity of the operators $\bm{\rho}$ and $\bm{p}_{\perp}$ (see Eqs. (\ref{F2})-(\ref{Ffinal})). Thus, the quantity $2\sqrt{\beta}\,\bm q$ can be interpreted as due to the quantum fluctuations of the particle around the classical rectilinear trajectory described by the equations $\bm \rho-x(\bm \rho-\bm \rho')=0$ and $T-x(T-T')=0$. We also note that Eq. (\ref{D0final}) is in agreement with the corresponding results obtained in Ref. \cite{LM95A} for the case of  spherically symmetric potential and in Ref. \cite{LMS00} for an arbitrary localized potential under the assumption $|\bm\rho-\bm\rho'|\ll|T-T'|$.

\subsection{The case of nonvanishing laser field}
In the case of the combined atomic and plane-wave background fields, the Green's function
$D^{(0)}(T,\,\bm\rho|\, T',\,\bm\rho';\,\epsilon)$ of the Klein-Gordon equation reads
\begin{equation}\label{KGVL}
D^{(0)}(T,\,\bm\rho|\, T',\,\bm\rho';\,\epsilon)=-i\int_0^\infty ds\,\exp\Big[-is\Big(2\epsilon V(\bm\rho,\,T)+H+m^2\Big)\Big]\delta(T-T')\delta(\bm\rho-\bm\rho'),
\end{equation}
where $H=2\epsilon p_T+[\bm p_\perp-\bm{\mathcal{A}}(T)]^2$. First, it is convenient to disentangle the operator $\exp(-isH)$ according to the exact relation (see Ref. \cite{Baier_1975_b}),
\begin{equation}
 \mbox{e}^{-isH}=\exp\Big[-is\int_0^1dx\Big(\bm p_\perp-\bm{\mathcal{A}}(T-2\epsilon s x)\Big)^2 \,\Big]\,\mbox{e}^{-2is\epsilon p_T}.
\end{equation}
Now, after expanding the square in the exponent of this expression, two terms arise containing the operator $\bm{p}_{\perp}$, one linearly and the other one quadratically, which commute with each other. The action of the operators with the exponent proportional to $p_T$ and $\bm{p}_{\perp}$ (to $\bm{p}_{\perp}^2$) on the $\delta$-functions in Eq. (\ref{KGVL}) can be inferred from Eqs. (\ref{exp_1}), (\ref{exp_1_p}), and (\ref{exp_p2}). The final result is
\begin{equation}\label{H1L}
\mbox{e}^{-isH}\delta(T-T')\delta(\bm\rho-\bm\rho')=-\frac{i}{4\pi s}\delta(T-T'-2\epsilon s)\exp\left[i\frac{(\bm\rho-\bm\rho')^2}{4s}+is(\bm f^2-g^2)+i(\bm\rho-\bm\rho')\cdot\bm f\right],
\end{equation}
where
\begin{align}
\bm f=\int_0^1dx\,\bm{\mathcal{A}}(T-x(T-T'))\,,&& g^2=\int_0^1dx\,\bm{\mathcal{A}}^2(T-x(T-T')).
\end{align}
Analogously to the previous case of pure atomic field, we disentangle now the terms proportional to $V(\bm\rho,\,T)$ and to $H$ in the exponential operator in Eq. (\ref{KGVL}) by introducing the operator $L(s)$, which is defined as
\begin{equation}
L(s)=\mbox{e}^{-is[2\epsilon V(\bm\rho,\,T)+H]}\mbox{e}^{isH}.
\end{equation}
By differentiating this operator with respect to $s$, and by employing again Eqs. (\ref{exp_1})-(\ref{exp_1_p}), it can be expressed in the form
\begin{equation}\label{FL}
L(s)= \exp\left[-2is\epsilon\int_0^1 V\Big(\bm\rho-2 sx\bm p_\perp+2sx\int_0^1dx'\bm{\mathcal{A}}(T-2s\epsilon xx'),\,T-2\epsilon sx\Big)\,dx\right],
\end{equation}
where, as in the previous section, we neglected the commutators among the operators at different values of $s$, as we work in the leading-order of the quasiclassical approximation.

Now, the last task is to apply the operator $L(s)$ to the function in Eq. (\ref{H1L}). As in the previous section, the main difficulty is the presence of the operator $\bm{p}_{\perp}$ under the integral in $x$ in Eq. (\ref{FL}), which can be overcome again referring to the fact that in a localized potential quantum fluctuations around the classical trajectory can be neglected unless $x\approx x_0=T/(T-T')$, with $0<x_0<1$. By repeating analogous steps as in the previous section, we obtain the following final result for the Green's function $D^{(0)}(T,\,\bm\rho|\, T',\,\bm\rho';\,\epsilon)$ (see also \cite{DPM2012}):
\begin{eqnarray}\label{D0finalL}
&&  D^{(0)}(T,\,\bm\rho|\, T',\,\bm\rho';\,\epsilon)=\frac{i\theta(s_0)}{4\pi^2|T-T'|}
\exp\Big[is_0(\bm f^2-g^2)+i(\bm\rho-\bm\rho')\cdot\bm f
\nonumber\\
&&+i\frac{(\bm\rho-\bm\rho')^2}{4s_0}-im^2s_0\Big]\int d\bm{q}\,\exp\left[iq^2-2is_0\epsilon\int_0^1 V(\bm \rho_x,T-x(T-T'))\,dx\right]\,,
\end{eqnarray}
where
\begin{equation}
\bm \rho_x=\bm \rho-x(\bm \rho-\bm \rho')+2\beta \tilde{\bm{f}}+2\sqrt{\beta}\,\bm q,
\end{equation}
with
\begin{equation}
\tilde{\bm{f}}=\int_0^1dy\,[\bm{\mathcal{A}}(Ty)-\bm{\mathcal{A}}(T'y)].
\end{equation}
The expression of $\bm{\rho}_x$ indicates that, as expected, the presence of the laser field alters the rectilinear trajectory in the transverse plane. Moreover, by performing the calculations, one first obtains that the term containing the laser field in the expression of $\bm{\rho}_x$ is proportional to $x(1-x)s_0$, which, however, can be approximated as $\beta=x_0(1-x_0)s_0$. In fact, we have assumed that the laser angular frequency $\omega_0$ is such that
$\omega_0\epsilon\ll m^2$. Thus, in the formation region of a typical QED process as the Bethe-Heitler one, where $|T-T'|\sim (\epsilon/m)\lambda_C$, the laser field is practically constant (see Ref. \cite{DPM2012} for additional details). Moreover, the alteration $\Delta\bm{\rho}$ of the trajectory in the transverse plane due to the laser field is about $E/E_c$ times smaller than $|T-T'|\sim (\epsilon/m)\lambda_C$, such that it can be neglected anyway in the expression of the atomic potential unless $x\approx x_0$ and $0<x_0<1$.

At this point, the Green's function of the Dirac equation can be represented in the form
\begin{equation}\label{FGD}
G(x,\,x')=\langle x|\frac{1}{\hat{\cal P}-m+i0}|x'\rangle=(\hat{\cal P}+m)D(x,\,x'),
\end{equation}
In this expression, we have introduced the propagator of the ``squared'' Dirac equation
\begin{equation}
D(x,\,x')=\langle x|\frac{1}{\hat{\cal P}^2-m^2+i0}|x'\rangle,
\end{equation}
and we have employed the notation $\hat{\cal P}=\gamma^\mu{\cal P}_\mu$, where $\gamma^\mu$ are the Dirac matrices and where ${\cal P}_\mu=i\partial_\mu-eA_\mu(x)$, with $eA^{\mu}(x)=(V(\bm{\rho},T),\bm{\mathcal{A}}(T),0)$. In the field configuration at hand, it is
\begin{equation}
{\hat{\cal P}}^2={\cal P}^2+i\bm\alpha\cdot \bm\nabla V(\bm \rho,\, T)+\frac{i}{2}\gamma_+\,
\bm \gamma\cdot\partial_T\bm{\mathcal{A}}(T)
\end{equation}
where
\begin{equation}
{\cal P}^2=-2\epsilon p_T-2\epsilon V(\bm\rho,\,T)-(\bm p_\perp-\bm{\mathcal{A}}(T))^2,
\end{equation}
with $\bm\nabla=\partial_{\bm\rho}+\bm z \partial_T$ and $\gamma_+=\gamma^0+\gamma^3$. It is convenient to introduce the effective potential $\tilde V(\bm{\rho},T)=V(\bm{\rho},T)+\delta V(\bm{\rho},T)$, where
\begin{equation}
\delta V(\bm{\rho},T)=-\frac{i}{2\epsilon}\left[\bm\alpha\cdot \bm\nabla V(\bm \rho,\, T)+\frac{1}{2}\gamma_+\,
\bm \gamma\cdot\partial_T\bm{\mathcal{A}}(T)\right]\,.
\end{equation}
It is important to note that, though $\delta V(\bm{\rho},T)$ is small in comparison with $V(\bm{\rho},T)$, it has another matrix structure and it should be taken into account in order to obtain consistent results. The corresponding correction in the propagator of the squared Dirac equation can be obtained by employing Eq. (\ref{D0finalL})
with the replacement $V(\bm{\rho},T)\rightarrow\tilde V(\bm{\rho},T)$ and then by performing the expansion with respect to $\delta V(\bm{\rho},T)$. The final result is
\begin{equation}\label{DfinalL}
D(T,\,\bm\rho|\, T',\,\bm\rho';\,\epsilon)=\left\{1-\frac{i}{2\epsilon}\bm\alpha\cdot\left(\partial_{\bm\rho}+
\partial_{\bm\rho'}\right)-\frac{\gamma_+}{4\epsilon}\bm\gamma\cdot[\bm{\mathcal{A}}(T)-\bm{\mathcal{A}}(T')]\right\} D^{(0)}(T,\,\bm\rho|\, T',\,\bm\rho';\,\epsilon)\,.
\end{equation}
This expression allows then to obtain the Green's function of the Dirac equation from the relation (see Eq. (\ref{FGD}))
\begin{equation}\label{GfinalL}
G(T,\,\bm\rho|\, T',\,\bm\rho';\,\epsilon)=\Big\{\gamma_-\epsilon+\frac{i}{2}\gamma_+\partial_T-\gamma^0V(\bm \rho,\, T)
+\bm{\gamma}_{\perp}\cdot[i\partial_{\bm{\rho}}+\bm{\mathcal{A}}(T)]+m\Big\}D(T,\,\bm\rho|\, T',\,\bm\rho';\,\epsilon),
\end{equation}
where $\gamma_-=\gamma^0-\gamma^3$ and $\bm{\gamma}_{\perp}=(\gamma^1,\gamma^2)$. We point out that this expression is exact with respect to the strength of the laser field and of the atomic potential.

\section{High-energy $e^+e^-$ photoproduction cross section}
By employing the Green's function in Eq. (\ref{GfinalL}), we have evaluated in \cite{DPM2012} the modification to the total Bethe-Heitler pair-production cross section due to the presence of a plane-wave field. We would like to present here some details of the derivation. Due to the optical theorem, the total cross section $\sigma$ of  $e^+e^-$ photoproduction can be expressed in terms of the forward Delbr\"uck scattering amplitude $M_D$ (coherent scattering of a photon in an external field via a virtual electron-positron pair),
\begin{eqnarray}\label{s1}
&& \sigma=\frac{1}{\omega}\mbox{Im}M_D\,,\nonumber\\
&& M_D=2i\alpha\int d\epsilon\,dT\,dT'd\bm\rho\,d\bm\rho'\,\mbox{Sp}\left\{\hat e^*\, G(T,\,\bm\rho|\, T',\,\bm\rho';\,\epsilon)\hat e\, G(T',\,\bm\rho'|\, T,\,\bm\rho;\,\epsilon-\omega)\right\}\,,
\end{eqnarray}
where $\hat e=-\bm\gamma\cdot\bm e$, and $e^{\mu}=(0,\bm e)$ is the polarization four-vector of the photon, with energy $\omega\gg m$. Following Ref. \cite{LM95A}, we rewrite $M_D$ as
\begin{eqnarray}\label{s2}
&& M_D=i\alpha\int_0^\omega d\epsilon \int dT\,dT'd\bm\rho\,d\bm\rho'\,\theta(T-T')
\mbox{Sp}\Big\{[(2\bm e^*\cdot{\bm{\mathcal{P}}}_\perp-\hat e^*\hat k) \, D(T,\,\bm\rho|\, T',\,\bm\rho';\,\epsilon)]\nonumber\\
&&\times [(2\bm e\cdot{\bm{\mathcal{P}}}'_\perp+\hat e\hat k) \, D(T',\,\bm\rho'|\, T,\,\bm\rho;\,\epsilon-\omega)]\Big\}\,,
\end{eqnarray}
where ${\bm{\mathcal{P}}}_\perp=  -i\partial_{\bm\rho}-{\bm{\mathcal{A}}}(T)$ and ${\bm{\mathcal{P}}}'_\perp=  -i\partial_{\bm\rho'}-{\bm{\mathcal{A}}}(T')$. Here, the $\theta$ function in $D^{(0)}(T,\,\bm\rho|\, T',\,\bm\rho';\,\epsilon)$ is taken into account (see Eq. (\ref{D0finalL})). Substituting Eq. (\ref{DfinalL}) and taking the trace, we obtain
\begin{eqnarray}\label{MD1}
&& M_D=4i\alpha\int_0^\omega d\epsilon \int dT\,dT'd\bm\rho\,d\bm\rho'\,\theta(T-T')
\Big\{4[(\bm e^*\cdot{\bm{\mathcal{P}}}_\perp) D_+^{(0)}][(\bm e\cdot{\bm{\mathcal{P}}}'_\perp)D_-^{(0)}] \,\nonumber\\
&&-\frac{\omega^2}{\epsilon(\omega-\epsilon)}\Big[[(\bm e^*\cdot{\bm{\mathcal{P}}}_\perp) D_+^{(0)}][(\bm e\cdot{\bm{\mathcal{P}}}'_\perp)D_-^{(0)}]-
 [(\bm e\cdot{\bm{\mathcal{P}}}_\perp) D_+^{(0)}][(\bm e^*\cdot{\bm{\mathcal{P}}}'_\perp)D_-^{(0)}]\nonumber\\
&&+\frac{1}{2}[({\bm{\mathcal{P}}}_\perp-{\bm{\mathcal{P}}}^{'*}_\perp) D_+^{(0)}]\cdot[({\bm{\mathcal{P}}}_\perp'-{\bm{\mathcal{P}}}^{*}_\perp) D_-^{(0)}]
 \Big\}\,,
\end{eqnarray}
where $D_+^{(0)}=D^{(0)}(T,\,\bm\rho|\, T',\,\bm\rho';\,\epsilon)$, $D_-^{(0)}=D^{(0)}(T',\,\bm\rho'|\, T,\,\bm\rho;\,\epsilon-\omega)$, ${\bm{\mathcal{P}}}^*_\perp=  i\partial_{\bm\rho}-{\bm{\mathcal{A}}}(T)$ and ${\bm{\mathcal{P}}}^{'*}_\perp=i\partial_{\bm\rho'}-{\bm{\mathcal{A}}}(T')$.

Below, we restrict ourselves by the case of incoming unpolarized photon and the expression Eq. (\ref{MD1}) essentially simplified,
\begin{eqnarray}\label{MD2}
&& M_D=4i\alpha\int_0^\omega d\epsilon \int dT\,dT'd\bm\rho\,d\bm\rho'\,\theta(T-T')
\Big\{2[{\bm{\mathcal{P}}}_\perp D_+^{(0)}]\cdot[{\bm{\mathcal{P}}}'_\perp D_-^{(0)}] \,\nonumber\\
&&-\frac{\omega^2}{2\epsilon(\omega-\epsilon)}
[({\bm{\mathcal{P}}}_\perp-{\bm{\mathcal{P}}}^{'*}_\perp) D_+^{(0)}]\cdot[({\bm{\mathcal{P}}}_\perp'-{\bm{\mathcal{P}}}^{*}_\perp) D_-^{(0)}]\Big\}\,.
\end{eqnarray}
Since we are interested in investigating the influence of the laser field on the cross section of $e^+e^-$ high-energy photoproduction in the electric field of a heavy atom, we assume that in the integrand in Eq. (\ref{MD2}) its value at  $\bm{\mathcal{A}}(T)=\bm{0}$ has been subtracted. Let us now represent the amplitude $M_D$ as a sum of the Born term $M_B$ (the contribution to $M_D$ leading in $Z\alpha$ and proportional to $(Z\alpha)^2$) and the Coulomb corrections $M_C$.

First, we determine the quantity $M_C$. The main contribution to the Coulomb corrections is given by the region of integration where $T$ is positive and $T'$ negative and where $\rho,\,\rho'\sim \lambda_C\ll T,\,|T'|$. Since the atomic screening radius $r_c$ can be assumed to be much larger than $\lambda_C$ (in the Thomas-Fermi model it is $r_c \sim \lambda_C/(\alpha Z^{1/3})$), we can replace the potential $V(\bm r)$ by the Coulomb potential $V_C(r)=-Z\alpha/r$. For the Coulomb potential at $T,\,|T'|\gg \rho,\rho'\sim \lambda_C$ we have
\begin{equation}
 \int_0^1 V_C(\bm{\rho}_x,T-x(T-T'))dx=\frac{2Z\alpha}{T-T'}\ln\left[\frac{1}{\sqrt{-TT'}}\left|\sqrt{\beta}\bm q+\beta \tilde{\bm{f}}+\frac{T\bm\rho'-T'\bm\rho}{2(T-T')}\right|\right]\,.
\end{equation}
Then, we exploit the relation $\omega_0T\sim \omega_0\omega/m^2\ll 1$, and expand $\bm{\mathcal{A}}(T)$ so that
$\bm{\mathcal{A}}(T)=\bm{\mathcal{A}}(0)+\bm{\mathcal{E}}T$, where  the quantity $\bm{\mathcal{E}}=\partial \bm{\mathcal{A}}/\partial T$ at $T=0$ depends on some constant phase. The final result should be averaged over this phase. Due to gauge invariance, the amplitude $M_D$ is independent of $\bm{\mathcal{A}}(0)$. Since there are no external vectors in the expression (\ref{MD2}), the amplitude  $M_D$ depends on $\bm{\mathcal{E}}$ via $\bm{\mathcal{E}}^2$. Note that, for circular polarization of the laser field, $\bm{\mathcal{E}}^2$ is independent of the phase of the field. For the contribution $M_C$ we have,
\begin{eqnarray}\label{MC1}
&& M_C=-\frac{i\alpha}{2\pi^4}\int_0^\omega d\epsilon \int_0^\infty\!\!\int_0^\infty \frac{dT\,dT'}{(T+T')^2}\int\!\!\int d\bm\rho\,d\bm\rho'\,\int\!\!\int d\bm q_1\,d\bm q_2\,\nonumber\\
&&\times\left[\left(\frac{|\bm F_2|}{|\bm F_1|}\right)^{2iZ\alpha}-1+2(Z\alpha)^2\ln^2\left(\frac{|\bm F_2|}{|\bm F_1|}\right)\right]\nonumber\\
&&\times\exp\Bigg\{i\Bigg[ q_1^2+q_2^2+\frac{\omega(\bm\rho-\bm\rho')^2}{2(T+T')}-\frac{\omega(T+T')}{2\epsilon (\omega-\epsilon)}\left(m^2+\frac{1}{12}\bm f_0^2\right)\Bigg]\Bigg\}\nonumber\\
 &&\times\Bigg\{\left[\frac{T'\bm q_1}{(T+T')\sqrt{\beta_1}}-\frac{\epsilon(\bm\rho-\bm\rho')}{T+T'}+\frac{\bm f_0}{2}\right]\cdot
\left[\frac{T\bm q_2}{(T+T')\sqrt{\beta_2}}+\frac{(\omega-\epsilon)(\bm\rho-\bm\rho')}{T+T'}-\frac{\bm f_0}{2}\right]\,\nonumber\\
&&-\frac{\omega^2}{4\epsilon(\omega-\epsilon)}
\left[\frac{\bm q_1}{\sqrt{\beta_1}}+\bm f_0\right]\cdot\left[\frac{\bm q_2}{\sqrt{\beta_2}}-\bm f_0\right]\Bigg\}\,,
\end{eqnarray}
where we have made the replacement $T'\rightarrow -T'$ and omit the $\theta$ function and where the following notations are used
\begin{eqnarray}
&&\bm F_1=\sqrt{\beta_1}\bm q_1+\frac{1}{2}\beta_1\bm f_0+\frac{T'\bm\rho+T\bm\rho'}{2(T+T')}\,,\quad
\bm F_2=\sqrt{\beta_2}\bm q_2-\frac{1}{2}\beta_2\bm f_0+\frac{T'\bm\rho+T\bm\rho'}{2(T+T')}\,,\nonumber\\
&&\bm f_0=(T+T')\bm{\mathcal{E}}\,,\quad \beta_1=\frac{TT'}{2\epsilon(T+T')}\,,\quad \beta_2=\frac{TT'}{2(\omega-\epsilon)(T+T')}\,.
\end{eqnarray}
Also, in the above expression of $M_C$ the first two terms in the expansion at $Z\alpha\ll 1$ have been subtracted: the first term (the unity in the second line in Eq. (\ref{MC1})) ensures that there is no pair production at $Z\to 0$; the second term (proportional to $(Z\alpha)^2$ in the second line in Eq. (\ref{MC1})) corresponds to the Born contribution. The corresponding integral, with the opposite sign, is the mentioned Born term $M_B$, which will be analyzed below. The calculation of the eleven-fold integral in Eq. (\ref{MC1}) is a non-trivial task and we present here some details of this calculation because they may be useful in the consideration of other processes with the use of the Green's function obtained in our work. Then, we pass from the variables $\bm q_1$ and $\bm q_2$ to the variables $\bm Q_1=\sqrt{\beta_1}\bm q_1 +(T'\bm\rho+T\bm\rho')/2(T+T')$ and
$\bm Q_2=\sqrt{\beta_2}\bm q_2 +(T'\bm\rho+T\bm\rho')/2(T+T')$ and take the integral over $\bm\rho$ and $\bm\rho'$,
\begin{eqnarray}\label{MC2}
&& M_C=\frac{8i\alpha}{\pi^2}\int_0^\omega d\epsilon\,a \int_0^\infty\!\!\int_0^\infty \frac{dT\,dT'}{TT'}\,\int\!\!\int d\bm Q_1\,d\bm Q_2\,\nonumber\\
&&\times\left[\left(\frac{|\bm Q_2-\beta_2\bm f_0/2|}{|\bm Q_1+\beta_1\bm f_0/2|}\right)^{2iZ\alpha}-1+2(Z\alpha)^2\ln^2\left(
\frac{|\bm Q_2-\beta_2\bm f_0/2|}{|\bm Q_1+\beta_1\bm f_0/2|}\right)\right]\nonumber\\
&&\times\exp\Bigg\{i\Bigg[\frac{2\omega a(T+T')(\bm Q_1-\bm Q_2)^2}{TT'}-\frac{T+T'}{2\omega a}\left(m^2+\frac{1}{12}\bm f_0^2\right)\Bigg]\Bigg\}\nonumber\\
 &&\times\Bigg\{\left(\frac{1}{a}-1\right)\frac{\bm f_0^2}{4}+a\omega^2\left[\frac{(T+T')^2}{TT'}-4 a\right]\frac{(\bm Q_1-\bm Q_2)^2}{TT'}\nonumber\\
&&+\frac{\omega(T+T')}{TT'} (1-a)(\bm Q_1-\bm Q_2,\,\bm f_0)-i\frac{\omega(T+T')}{2TT'}\Bigg\}\,,
\end{eqnarray}
where $a=\epsilon(\omega-\epsilon)/\omega^2$. Then, we make the change of the variables, $\bm Q_1=\tilde {\bm Q}-\bm Q/2+(\beta_2-\beta_1)\bm f_0/4$ and
$\bm Q_2=\tilde {\bm Q}+\bm Q/2+(\beta_2-\beta_1)\bm f_0/4$. The integration over $\tilde{\bm Q}$ gives
\begin{eqnarray}\label{MC3}
&& M_C=\frac{2i\alpha}{\pi^2}G(Z\alpha)\int_0^\omega d\epsilon\,a \int_0^\infty\!\!\int_0^\infty \frac{dT\,dT'}{TT'}\,\int\,d\bm Q
\,\left[\bm Q-\frac{1}{2}(\beta_1+\beta_2)\bm f_0\right]^2\nonumber\\
&&\times\exp\Bigg\{i\Bigg[\frac{2\omega a(T+T')\bm Q^2}{TT'}-\frac{(T+T')}{2\omega a}\left(m^2+\frac{1}{12}\bm f_0^2\right)\Bigg]\Bigg\}\nonumber\\
 &&\times\Bigg\{\left(\frac{1}{a}-1\right)\frac{\bm f_0^2}{4}+a\omega^2\left[\frac{(T+T')^2}{TT'}-4 a\right]\frac{\bm Q^2}{TT'}\nonumber\\
&& -\frac{\omega(T+T')}{TT'} (1-a)(\bm Q\cdot\bm f_0) -i\frac{\omega(T+T')}{2TT'}\Bigg\}\,,
\end{eqnarray}
where
\begin{equation}
\label{G}
\begin{split}
G(Z\alpha)&=\int d\bm q \left[\left(\frac{|\bm q-\bm s|}{|\bm q+\bm s|}\right)^{2iZ\alpha}-1+2(Z\alpha)^2\ln^2\left(
\frac{|\bm q-\bm s|}{|\bm q+\bm s|}\right)\right]\\
&=8\pi(Z\alpha)^2\,[\mbox{Re}\psi(1+iZ\alpha)+C]=8\pi(Z\alpha)^2 f(Z\alpha)\,,
\end{split}
\end{equation}
with $\bm s$ being an arbitrary two-dimensional unit vector, with $\psi(t)=d\ln\Gamma(t)/dt$, and with $C=0.577\dots$ being the Euler constant. The function $G(Z\alpha)$ is derived in the Appendix B of Ref. \cite{LMS2004}. Then, we  make change of the variables, $T=2(\omega a/m^2)u\tau$, $T'=2(\omega a/m^2)(1-u)\tau$,
$\epsilon=\omega x$, and take the integrals over $\bm Q$ and over $u$. We obtain,
\begin{eqnarray}\label{MCfinal}
&& M_C=-\frac{\alpha\omega(Z\alpha)^2f(Z\alpha)}{m^2}\Phi(\chi)\,,\nonumber\\
&&\Phi (\chi)=16\int_0^1 dx \int_0^\infty d\tau \exp\left\{-i\tau\left[1+\frac{4}{3}\chi^2a^2\tau^2\right]\right\}\nonumber\\
&&\times\left[-\frac{1}{4}+\frac{1}{3}a+\frac{4}{3}i\chi^2\tau^3 a^2\left(1-\frac{11}{10}a\right)\
+\frac{8}{15}\chi^4\tau^6a^4(1-a)\right]\,.
\end{eqnarray}
We note that $\bm{\mathcal{E}}=-2e\bm{E}$, where $\bm{E}$ is the electric field of the laser wave, and that the integral over $\tau$ can be expressed via the Scorer functions. The asymptotic forms of the function $\Phi (\chi)$ at small values of $\chi$ reads
\begin{eqnarray}\label{as}
&&\Phi (\chi)=\frac{28i}{9}\left(1+\frac{1584}{1225}\chi^2\right).
\end{eqnarray}
The asymptotics of the function $\Phi (\chi)$ at large $\chi$ can also be easily calculated and it is proportional to $\chi^{-2/3}$. However, this asymptotics is applicable only at very large values of $\chi$.

Let us consider now the Born term $M_B$. It is clear that the integral in $\bm{q}$ of the last term in the integrand of the function $G(Z\alpha)$ (see Eq. (\ref{G})) is logarithmically divergent at large $q$'s. Therefore, the calculation of the Born term is a more delicate task and we follow the method described in detail in Refs. \cite{LM95A,LMS00}. Since we want to have as large as possible values of the parameter $\chi$, we consider the case when $\omega/m^2\gg r_c$. We pass from the variables $\bm\rho$ and $\bm\rho'$ to the variables $\bm\rho_-=\bm\rho-\bm\rho'$ and $\bm\rho_+=(T'\bm\rho+T\bm\rho')/(T+T')$. Then, we divide the region of integration over $\bm{\rho}_+$ into two: between $0$ and $\kappa$, and between $\kappa $ and $\infty$, where $\lambda_C\ll \kappa  \ll r_c$. In the first region, we can replace the potential $V(\bm{r})$ by the Coulomb one. Thus, we obtain the contribution of the first region to the Born term,
\begin{eqnarray}\label{MB1}
&& M_{B1}=\frac{\alpha\omega(Z\alpha)^2}{m^2}\left\{\Phi (\chi)\left[\ln\left(\frac{m\kappa  }{2}\right)+C+\frac{41}{42}\right] +\Psi(\chi)\right\}\,,
\end{eqnarray}
where $\Psi(\chi)=\Psi_{1}(\chi)+\Psi_{2}(\chi)$, with
\begin{eqnarray}\label{phi11}
&&\Psi_{1}(\chi) =\int_0^1 dx \int_0^\infty d\tau \exp\left\{-i\tau\left[1+\frac{4}{3}\chi^2a^2\tau^2\right]\right\}
\Bigg\{8\left[-\log(\tau)-C-i\frac{\pi}{2}\right]\nonumber\\
&&\times\left[-\frac{1}{4}+\frac{1}{3}a+\frac{4}{3}i\chi^2\tau^3 a^2\left(1-\frac{11}{10}a\right)\
+\frac{8}{15}\chi^4\tau^6a^4(1-a)\right]-\frac{2}{21}+\frac{4a}{7}\nonumber\\
&&+i\chi^2\tau^3a^2\left[\frac{44}{7}-\frac{1448}{175}a\right]
 +\frac{3616}{525}\chi^4\tau^6a^4(1-a)\Bigg\}
\end{eqnarray}
and
\begin{eqnarray}\label{phi12}
&&\Psi_{2} =-\frac{4i}{\pi}\int_0^1dy\int_0^1 dx \int_0^\infty d\tau \exp\left\{-i\tau\left[1+\frac{4}{3}\chi^2a^2\tau^2\right]\right\}\nonumber\\
&&\times\int d\bm q \ln\left[\frac{q}{|\bm q+\bm F|}\right]\,\mbox{e}^{iq ^2}\,(\bm q+\bm F)^2\,
\big[a (2y-1)^2q^2+(1-a)(\bm q+2\bm F)^2-i\big]\,,
\end{eqnarray}
with $\bm F=2\chi a\sqrt{y(1-y)}\tau^{3/2}\bm s$ and $\bm s$ being arbitrary unit vector.

In the second region we use the relation $|\sqrt{\beta}\bm q +\beta \tilde{\bm{f}}|\ll \rho_+$ and expand the potential $V(\bm \rho_x,T-x(T-T'))$ in the Green function Eq. (\ref{D0finalL}). As a result, the part of the phase in the integrand of the matrix element, depending on the potential, has the form
\begin{align}
\varphi&=-2\left[\sqrt{\beta_1}\bm q_1-\sqrt{\beta_2}\bm q_2+(\beta_1+\beta_2)\tilde{\bm{f}}\right]
\cdot\frac{\partial\delta(\rho_+)}{\partial\bm\rho_+}\,,\\
\delta(\rho)&=\int_{-\infty}^{\infty}V(\bm\rho,\,x)dx\,.
\end{align}
Then, we expand the integrand over $\varphi$, keep the term proportional to $\varphi^2$ and integrate it over $\bm\rho_-$, $\bm q_1$ and $\bm q_2$. Finally, the contribution to the Born amplitude from the second region reads
\begin{eqnarray}\label{MB2}
&& M_{B2}=\frac{\alpha\omega(Z\alpha)^2}{m^2}\Phi (\chi)\left[\frac{1}{4(Z\alpha)^2}\int_{\kappa }^\infty \rho\left(\frac{\partial\delta}{\partial\rho}\right)^2\,d\rho\right]\,,
\end{eqnarray}
Then, we use the result of  Ref. \cite{LM95A} for the integral over $\rho$, obtained within Thomas-Fermi model, and arrive to the final expression of the total amplitude:
\begin{equation}\label{MDfinal}
M_{D}=M_{B1}+M_{B2}+M_C=\frac{\alpha\omega(Z\alpha)^2}{m^2}\left\{\Phi (\chi)\left[\ln(183 Z^{-1/3})-\frac{1}{42}-f(Z\alpha)\right]+\Psi(\chi)\right\}\,.
\end{equation}
By applying the optical theorem, the cross section $\sigma$ is finally given by (see Eq. (\ref{s1}))
\begin{equation}
\sigma=\frac{1}{\omega}\text{Im}M_D=\sigma_0\left\{\text{Im}\Phi(\chi)\left[\ln(183 Z^{-1/3})-\frac{1}{42}-f(Z\alpha)\right]+\text{Im}\Psi(\chi)\right\},
\end{equation}
where $\sigma_0=\alpha(Z\alpha)^2/m^2$, see also footnote \cite{Footnote}. The asymptotic forms of the functions $\Phi(\chi)$ and $\Psi(\chi)$ at $\chi\ll 1$ are \cite{Footnote}
\begin{align}
\text{Im}\Phi(\chi)=\frac{28}{9}\Big(1+\frac{1584}{1225}\chi^2\Big), && \text{Im}\Psi(\chi)=-\frac{195088}{55125}\chi^2
\end{align}
and are in agreement with the corresponding formulas in \cite{Baier_b_1998}.

\section{Quasiclassical wave functions in combined atomic and plane-wave field}
In this section, we derive the quasiclassical wave functions which include exactly both the atomic and the plane-wave field, starting from the Green's functions obtained above. As in the previous section, it is instructive to first derive the corresponding relations in the case of vanishing laser field. These relations can already be found, e.g., in Ref. \cite{LMS00}.

\subsection{The case of vanishing laser field}

The Green's function $G(\bm{r}|\bm{r}';\epsilon)=[\gamma^0(\epsilon-V(\bm{r}))+i\bm{\gamma}\cdot\bm{\nabla}-m+i0]^{-1}\delta(\bm{r}-\bm{r}')$ of the Dirac equation in the presence of a potential $V(\bm{r})$ satisfies the relations
\begin{equation}
\label{G_G0}
G(\bm{r}|\bm{r}';\epsilon)=G_0(\bm{r}|\bm{r}';\epsilon)+\int d\bm{r}''G_0(\bm{r}|\bm{r}'';\epsilon)\gamma^0V(\bm{r}'')G(\bm{r}''|\bm{r}';\epsilon),
\end{equation}
where $G_0(\bm{r}|\bm{r}';\epsilon)$ is the corresponding Green's function of the Dirac equation in vacuum
\begin{equation}
G_0(\bm{r}|\bm{r}';\epsilon)=-\frac{1}{4\pi}(\gamma^0\epsilon+i\bm{\gamma}\cdot\bm{\nabla}+m)\frac{\mbox{e}^{ik|\bm{r}-\bm{r'}|}}{|\bm{r}-\bm{r'}|},
\end{equation}
where $k=\sqrt{\epsilon^2-m^2}$. Starting from the relation in Eq. (\ref{G_G0}) and assuming that $\epsilon>0$, it is easy to show that
\begin{equation}
\label{lim_G}
\lim_{r\to\infty}G(\bm{r}|\bm{r}';\epsilon)=-\frac{1}{4\pi}\frac{\mbox{e}^{ik r}}{r}\sum_{\lambda=1}^2 2\epsilon\, u_{p,\lambda}\bar{u}_{p,\lambda}\left[\mbox{e}^{-ik\bm{n}\cdot\bm{r}'}+\int d\bm{r}''\mbox{e}^{-ik\bm{n}\cdot\bm{r}''}\gamma^0V(\bm{r}'')G(\bm{r}''|\bm{r}';\epsilon)\right],
\end{equation}
where $p^{\mu}=(\epsilon,k\bm{n})$, with $\bm{n}=\bm{r}/r$, and where $u_{p,\lambda}$ is the positive-energy constant bi-spinor normalized as $\sum_{\lambda=1}^2 2\epsilon\, u_{p,\lambda}\bar{u}_{p,\lambda}=\hat{p}+m$ ($\bar{u}_{p,\lambda}=u^{\dag}_{p,\lambda}\gamma^0$) \cite{Landau_b_4_1982}. Now, the solution $U^{(\text{out})}_{p,\lambda}(\bm{r})$ of the Dirac equation in the potential $V(\bm{r})$ corresponding at large distances from the origin to a plane wave with momentum $k\bm{n}$ plus a converging spherical wave is given by
\begin{equation}
U^{(\text{out})}_{p,\lambda}(\bm{r})=\left[\mbox{e}^{ik\bm{n}\cdot\bm{r}}+\int d\bm{r}'\mbox{e}^{ik\bm{n}\cdot\bm{r}'}\tilde{G}(\bm{r}|\bm{r}';\epsilon)\gamma^0V(\bm{r}')\right]u_{p,\lambda},
\end{equation}
where $\tilde{G}(\bm{r}|\bm{r}';\epsilon)=[\gamma^0(\epsilon-V(\bm{r}))+i\bm{\gamma}\cdot\bm{\nabla}-m-i0]^{-1}\delta(\bm{r}-\bm{r}')$. Since $\bar{G}(\bm{r}|\bm{r}';\epsilon)=\tilde{G}(\bm{r}'|\bm{r};\epsilon)$, with $\bar{G}(\bm{r}|\bm{r}';\epsilon)=\gamma^0G^{\dag}(\bm{r}|\bm{r}';\epsilon)\gamma^0$, it can be shown that (see, e.g., Ref. \cite{LMS00})
\begin{equation}
\label{lim_G_r_p}
\lim_{r\to\infty}G(\bm{r}|\bm{r}';\epsilon)=-\frac{1}{4\pi}\frac{\mbox{e}^{ik r}}{r}\sum_{\lambda=1}^22\epsilon\,u_{p,\lambda}\bar{U}^{(\text{out})}_{p,\lambda}(\bm{r}'),
\end{equation}
from which the wave function $\bar{U}^{(\text{out})}_{p,\lambda_0}(\bm{r}')$ is obtained after projecting both sides onto $u_{p,\lambda_0}$.

In a completely analogous way, it can be show that
\begin{equation}
\label{lim_G_rp_p}
\lim_{r'\to\infty}G(\bm{r}|\bm{r}';\epsilon)=-\frac{1}{4\pi}\frac{\mbox{e}^{ik r'}}{r'}\sum_{\lambda=1}^22\epsilon\,U^{(\text{in})}_{p',\lambda}(\bm{r})\bar{u}_{p',\lambda},
\end{equation}
where $p^{\prime\,\mu}=(\epsilon,-k\bm{n}')$, with $\bm{n}'=\bm{r}'/r'$ and where $U^{(\text{in})}_{p',\lambda}(\bm{r})$ is the solution of the Dirac equation in the potential $V(\bm{r})$ corresponding at large distances from the origin to a plane wave with momentum $-k\bm{n}'$ plus a diverging spherical wave. The negative-energy states can also be obtained with this method, starting from the Green's function $G(\bm{r}|\bm{r}';-|\epsilon|)$. We present here only the final results (see also Ref. \cite{LMS00})
\begin{align}
\label{lim_G_r_m}
\lim_{r\to\infty}G(\bm{r}|\bm{r}';-|\epsilon|)&=\frac{1}{4\pi}\frac{\mbox{e}^{ik r}}{r}\sum_{\lambda=1}^2 2|\epsilon|\,v_{q,\lambda}\bar{V}^{(\text{out})}_{q,\lambda}(\bm{r}'),\\
\label{lim_G_rp_m}
\lim_{r'\to\infty}G(\bm{r}|\bm{r}';-|\epsilon|)&=\frac{1}{4\pi}\frac{\mbox{e}^{ik r'}}{r'}\sum_{\lambda=1}^2 2|\epsilon|\,V^{(\text{in})}_{q',\lambda}(\bm{r})\bar{v}_{q',\lambda},
\end{align}
where $q^{\mu}=(|\epsilon|,-k\bm{n})$, $q^{\prime\mu}=(|\epsilon|,k\bm{n}')$, and $v_{q,\lambda}$ is the negative-energy bi-spinor normalized as $\sum_{\lambda=1}^2 2|\epsilon|\,v_{q,\lambda}\bar{v}_{q,\lambda}=\hat{q}-m$ \cite{Landau_b_4_1982}.

\subsection{The case of non-vanishing laser field}
Before considering the case of combined atomic and plane-wave field, it is useful to first apply the method of the previous paragraph to the case in which only a plane wave is present. We point out that, although both the Green's function and the wave functions of a particle in a plane-wave field are known \cite{Landau_b_4_1982}, the relations obtained below are derived here for the first time. As we have pointed out in \cite{DPM2012}, the exact Green's function $D_w^{(0)}(T,\,\bm\rho|\, T',\,\bm\rho';\,\epsilon)$ ($G_w(T,\,\bm\rho|\, T',\,\bm\rho';\,\epsilon)$) of the Klein-Gordon (Dirac) equation in the presence of a plane wave can be obtained in the ultrarelativistic limit directly from Eq. (\ref{D0finalL}) (Eq. (\ref{GfinalL})) by setting $V(\bm{\rho},T)=0$. In the case of a scalar particle, it is
\begin{equation}
\begin{split}
&D_w^{(0)}(T,\,\bm\rho|\, T',\,\bm\rho';\,\epsilon)=-\frac{1}{4\pi|T-T'|}\theta\left(\frac{T-T'}{\epsilon}\right)\exp\left[\frac{i}{2}\frac{(\bm{\rho}-\bm{\rho}')^2}{T-T'}\epsilon-\frac{i}{2}\frac{m^2}{\epsilon}(T-T')\right]\\
&\quad\times\exp\left\{i\frac{\bm{\rho}-\bm{\rho}'}{T-T'}\cdot\int_{T'}^Td\tau \bm{\mathcal{A}}(\tau)+\frac{i}{2\epsilon}\left[\frac{1}{T-T'}\left(\int_{T'}^Td\tau \bm{\mathcal{A}}(\tau)\right)^2-\int_{T'}^Td\tau \bm{\mathcal{A}}^2(\tau)\right]\right\}.
\end{split}
\end{equation}
From this expression, it is straightforward to derive the four equalities
\begin{align}
\lim_{\substack{T\to\infty\\ \rho\to\infty}}D_w^{(0)}(T,\,\bm\rho|\, T',\,\bm\rho';\,\epsilon)&=-\frac{1}{4\pi}\frac{\mbox{e}^{i(p_{\perp}^2-m^2)T/2\epsilon}}{T}\phi_{w ;p}^{(\text{out})*}(T',\bm{\rho}') && \epsilon>0,\\
\lim_{\substack{T'\to-\infty\\ \rho'\to\infty}}D_w^{(0)}(T,\,\bm\rho|\, T',\,\bm\rho';\,\epsilon)&=-\frac{1}{4\pi}\frac{\mbox{e}^{i(p_{\perp}^{\prime\,2}-m^2)|T'|/2\epsilon}}{|T'|}\phi_{w;p'}^{(\text{in})}(T,\bm{\rho}) && \epsilon>0,\\
\lim_{\substack{T\to-\infty\\ \rho\to\infty}}D_w ^{(0)}(T,\,\bm\rho|\, T',\,\bm\rho';\,-|\epsilon|)&=-\frac{1}{4\pi}\frac{\mbox{e}^{i(q_{\perp}^2-m^2)|T|/2|\epsilon|}}{|T|}\phi_{w ;-q}^{(\text{out})*}(T',\bm{\rho}') && \epsilon<0,\\
\lim_{\substack{T'\to\infty\\ \rho'\to\infty}}D_w ^{(0)}(T,\,\bm\rho|\, T',\,\bm\rho';\,-|\epsilon|)&=-\frac{1}{4\pi}\frac{\mbox{e}^{i(q_{\perp}^{\prime\,2}-m^2)T'/2|\epsilon|}}{T'}\phi_{w ;-q'}^{(\text{in})}(T,\bm{\rho}) && \epsilon<0,
\end{align}
where
\begin{align}
p^{\mu}&=\epsilon\left(1,\frac{\bm{\rho}}{T},1-\frac{\rho^2}{2T^2}-\frac{m^2}{2\epsilon^2}\right), & q^{\mu}&=|\epsilon|\left(1,-\frac{\bm{\rho}}{|T|},1-\frac{\rho^2}{2T^2}-\frac{m^2}{2\epsilon^2}\right),\\
p^{\prime\mu}&=\epsilon\left(1,-\frac{\bm{\rho}'}{|T'|},1-\frac{\rho^{\prime\,2}}{2T^{\prime\,2}}-\frac{m^2}{2\epsilon^2}\right), & q^{\prime\mu}&=|\epsilon|\left(1,\frac{\bm{\rho}'}{T'},1-\frac{\rho^{\prime\,2}}{2T^{\prime\,2}}-\frac{m^2}{2\epsilon^2}\right),
\end{align}
and where $\phi_{w ;p}^{(\text{out}/\text{in})}(T,\bm{\rho})$ ($\phi_{w ;- p}^{(\text{out}/\text{in})}(T,\bm{\rho})$) are the positive-energy (negative-energy) scalar Volkov wave functions, which reduce to plane waves at $T\to\infty$ and $T\to-\infty$, respectively ($T\to-\infty$ and $T\to\infty$, respectively) \cite{Landau_b_4_1982}. The difference in the definition of in and out states for positive- and negative-energy states will be clear below when the case of combined atomic and laser field will be considered.

The spinor case can be worked out in a similar way and we report here only the final results
\begin{align}
\label{V_T_infty}
\lim_{\substack{T\to\infty\\ \rho\to\infty}}G_w (T,\,\bm\rho|\, T',\,\bm\rho';\,\epsilon)&=-\frac{1}{4\pi}\frac{\mbox{e}^{i(p_{\perp}^2-m^2)T/2\epsilon}}{T}\sum_{\lambda=1}^22\epsilon\,u_{p,\lambda}\bar{U}_{w ;p,\lambda}^{(\text{out})}(T',\bm{\rho}') && \epsilon>0,\\
\lim_{\substack{T'\to-\infty\\ \rho'\to\infty}}G_w (T,\,\bm\rho|\, T',\,\bm\rho';\,\epsilon)&=-\frac{1}{4\pi}\frac{\mbox{e}^{i(p_{\perp}^{\prime\,2}-m^2)|T'|/2\epsilon}}{|T'|}\sum_{\lambda=1}^2 2\epsilon\,U_{w ;p',\lambda}^{(\text{in})}(T,\bm{\rho})\bar{u}_{p',\lambda} && \epsilon>0,\\
\lim_{\substack{T\to-\infty\\ \rho\to\infty}}G_w (T,\,\bm\rho|\, T',\,\bm\rho';\,-|\epsilon|)&=\frac{1}{4\pi}\frac{\mbox{e}^{i(q_{\perp}^2-m^2)|T|/2|\epsilon|}}{|T|}\sum_{\lambda=1}^2 2|\epsilon|\,v_{q,\lambda}\bar{V}_{w ;q,\lambda}^{(\text{out})}(T',\bm{\rho}') && \epsilon<0,\\
\label{V_Tp_infty}
\lim_{\substack{T'\to\infty\\ \rho'\to\infty}}G_w (T,\,\bm\rho|\, T',\,\bm\rho';\,-|\epsilon|)&=\frac{1}{4\pi}\frac{\mbox{e}^{i(q_{\perp}^{\prime\,2}-m^2)T'/2|\epsilon|}}{T'}\sum_{\lambda=1}^22|\epsilon|\,V_{w ;q',\lambda}^{(\text{in})}(T,\bm{\rho})\bar{v}_{q',\lambda} && \epsilon<0,
\end{align}
valid for the positive-energy (negative-energy) spinor Volkov wave functions $U_{w ;p,\lambda}^{(\text{out})}(T,\rho)$ and $U_{w ;p,\lambda}^{(\text{in})}(T,\rho)$ ($V_{w ;p,\lambda}^{(\text{out})}(T,\rho)$ and $V_{w ;p,\lambda}^{(\text{in})}(T,\rho))$, which reduce to the corresponding free wave functions at $T\to\infty$ and $T\to-\infty$, respectively ($T\to-\infty$ and $T\to\infty$, respectively) \cite{Landau_b_4_1982}. By inspecting at the above results in the cases of atomic field (Eqs. (\ref{lim_G_r_p})-(\ref{lim_G_rp_m})) and of plane-wave field (Eqs.  (\ref{V_T_infty})-(\ref{V_Tp_infty})), one can easily deduce that the equalities (\ref{V_T_infty})-(\ref{V_Tp_infty}) are formally valid also in the case of combined atomic and plane-wave field once the index $w$ is removed and it is understood that the resulting quantities indicate the Green's functions and the wave functions in the combined-field case. In fact, it is clear that the Green's function $G(T,\,\bm\rho|\, T',\,\bm\rho';\,\epsilon)$ of the Dirac equation in the presence of the combined atomic and plane-wave field can be related to the Volkov Green's function $G_w (T,\,\bm\rho|\, T',\,\bm\rho';\,\epsilon)$ as (see also Eq. (\ref{G_G0}))
\begin{equation}
\begin{split}
G(T,\,\bm\rho|\, T',\,\bm\rho';\,\epsilon)=&G_w (T,\,\bm\rho|\, T',\,\bm\rho';\,\epsilon)\\
&+\int dT''d\bm{\rho}''G_w (T,\,\bm\rho|\, T'',\,\bm\rho'';\,\epsilon)\gamma^0V(\bm{\rho}'',T'')G(T'',\,\bm\rho''|\, T',\,\bm\rho';\,\epsilon)
\end{split}
\end{equation}
and that analogous relations hold for the corresponding wave functions. Thus, by performing the four different limits in the expression (\ref{GfinalL}) of the Green's function of the Dirac equation (see also Eqs. (\ref{D0finalL}) and (\ref{DfinalL})), one finally obtains the following spinor wave functions:
\begin{align}
\label{psi_pp}
\begin{split}
U_{p,\lambda}^{(\text{in})}(T,\bm{\rho})&=\exp\left\{-i\frac{m^2+p_{\perp}^2}{2\epsilon}T+i\bm{p}_{\perp}\cdot\bm{\rho}+i\frac{\bm{p}_{\perp}}{\epsilon}\cdot\int^T_{-\infty}d\tau\bm{\mathcal{A}}(\tau)-\frac{i}{2\epsilon}\int_{-\infty}^Td\tau\bm{\mathcal{A}}^2(\tau)\right\}\\
&\times\left[1-\frac{i}{2\epsilon}\bm{\alpha}\cdot\partial_{\bm{\rho}}-\frac{\gamma_+}{4\epsilon}\bm{\gamma}\cdot\bm{\mathcal{A}}(T)\right]u_{p,\lambda}\int \frac{d\bm{q}}{i\pi} \exp\left[iq^2-i\int_0^{\infty}d\tau\,V(\bm{\rho}_-,T-\tau) \right],
\end{split}\\
\begin{split}
U_{p,\lambda}^{(\text{out})}(T,\bm{\rho})&=-\exp\left\{-i\frac{m^2+p_{\perp}^2}{2\epsilon}T+i\bm{p}_{\perp}\cdot\bm{\rho}-i\frac{\bm{p}_{\perp}}{\epsilon}\cdot\int_T^{\infty}d\tau \bm{\mathcal{A}}(\tau)+\frac{i}{2\epsilon}\int_T^{\infty}d\tau\bm{\mathcal{A}}^2(\tau)\right\}\\
&\times\left[1-\frac{i}{2\epsilon}\bm{\alpha}\cdot\partial_{\bm{\rho}}-\frac{\gamma_+}{4\epsilon}\bm{\gamma}\cdot\bm{\mathcal{A}}(T)\right]u_{p,\lambda}\int \frac{d\bm{q}}{i\pi} \exp\left[-iq^2+i\int_0^{\infty}d\tau\,V(\bm{\rho}_+,T+\tau) \right],
\end{split}\\
\begin{split}
V_{p,\lambda}^{(\text{in})}(T,\bm{\rho})&=\exp\left\{i\frac{m^2+p_{\perp}^2}{2\epsilon}T-i\bm{p}_{\perp}\cdot\bm{\rho}-i\frac{\bm{p}_{\perp}}{\epsilon}\cdot\int_T^{\infty}d\tau\bm{\mathcal{A}}(\tau)-\frac{i}{2\epsilon}\int_T^{\infty}d\tau\bm{\mathcal{A}}^2(\tau)\right\}\\
&\times\left[1+\frac{i}{2\epsilon}\bm{\alpha}\cdot\partial_{\bm{\rho}}+\frac{\gamma_+}{4\epsilon}\bm{\gamma}\cdot\bm{\mathcal{A}}(T)\right]v_{p,\lambda}\int \frac{d\bm{q}}{i\pi} \exp\left[iq^2+i\int_0^{\infty}d\tau\,V(\tilde{\bm{\rho}}_-,T+\tau) \right],
\end{split}\\
\label{psi_mm}
\begin{split}
V_{p,\lambda}^{(\text{out})}(T,\bm{\rho})&=-\exp\left\{i\frac{m^2+p_{\perp}^2}{2\epsilon}T-i\bm{p}_{\perp}\cdot\bm{\rho}+i\frac{\bm{p}_{\perp}}{\epsilon}\cdot\int^T_{-\infty}d\tau\bm{\mathcal{A}}(\tau)+\frac{i}{2\epsilon}\int_{-\infty}^Td\tau\bm{\mathcal{A}}^2(\tau)\right\}\\
&\times\left[1+\frac{i}{2\epsilon}\bm{\alpha}\cdot\partial_{\bm{\rho}}+\frac{\gamma_+}{4\epsilon}\bm{\gamma}\cdot\bm{\mathcal{A}}(T)\right]v_{p,\lambda}\int \frac{d\bm{q}}{i\pi} \exp\left[-iq^2-i\int_0^{\infty}d\tau\,V(\tilde{\bm{\rho}}_+,T-\tau) \right],
\end{split}
\end{align}
where $\epsilon>0$ and where
\begin{align}
\bm{\rho}_{\pm}&=\bm{\rho}\pm\tau\frac{\bm{p}_{\perp}}{\epsilon}+\left[\sqrt{\frac{\mp 2T}{\epsilon}}\bm{q}+\frac{1}{\epsilon}\int_0^Tdy\bm{\mathcal{A}}(y)\right]\theta(\mp T),\\
\tilde{\bm{\rho}}_{\pm}&=\bm{\rho}\mp\tau\frac{\bm{p}_{\perp}}{\epsilon}+\left[\sqrt{\frac{\pm 2T}{\epsilon}}\bm{q}-\frac{1}{\epsilon}\int_0^Tdy\bm{\mathcal{A}}(y)\right]\theta(\pm T).
\end{align}
We point out that the notation of states as in/out corresponds to their behavior at large distances from the atomic center depending on if they stem from a plane wave and a diverging/converging spherical wave as in the case of pure atomic field. The  in and out states with positive energy (negative energy) for $T<0$ and $T>0$, respectively ($T>0$ and $T<0$, respectively) coincide with the eikonal wave functions. Also, by setting $\bm{\mathcal{A}}(T)=\bm{0}$, the wave functions in Eqs. (\ref{psi_pp})-(\ref{psi_mm}) reduce to those already obtained in \cite{LMS00} for $|T|\gg |\bm{\rho}|$. Finally, in the case of the Coulomb potential ($V_C(r)=-Z\alpha/r$) the integrals in $\bm{q}$ in Eqs. (\ref{psi_pp})-(\ref{psi_mm}) can be performed analytically. Apart from an inessential constant phase factor and by also including the dependence on the variable $\phi$, the resulting states are given by
\begin{align}
\label{psi_pp_C}
\begin{split}
U_{C;p,\lambda}^{(\text{in})}(x)&=C_+\exp\left[-i(pX_+)-\frac{i}{2\epsilon}\int_0^Td\tau\bm{\mathcal{A}}^2(\tau)\right]\\
&\times\left[1-\frac{i}{2\epsilon}\bm{\alpha}\cdot\partial_{\bm{\rho}}-\frac{\gamma_+}{4\epsilon}\bm{\gamma}\cdot\bm{\mathcal{A}}(T)\right]u_{p,\lambda}F(iZ\alpha,1,i(pR_+-\bm{p}\cdot\bm{R}_+)),\quad T>0
\end{split}\\
\begin{split}
U_{C;p,\lambda}^{(\text{out})}(x)&=C_+^*\exp\left[-i(pX_+)-\frac{i}{2\epsilon}\int_0^Td\tau\bm{\mathcal{A}}^2(\tau)\right]\\
&\times\left[1-\frac{i}{2\epsilon}\bm{\alpha}\cdot\partial_{\bm{\rho}}-\frac{\gamma_+}{4\epsilon}\bm{\gamma}\cdot\bm{\mathcal{A}}(T)\right]u_{p,\lambda}F(-iZ\alpha,1,-i(pR_++\bm{p}\cdot\bm{R}_+)),\quad T<0
\end{split}\\
\begin{split}
V_{C;p,\lambda}^{(\text{in})}(x)&=C_-\exp\left[i(pX_-)+\frac{i}{2\epsilon}\int_0^Td\tau\bm{\mathcal{A}}^2(\tau)\right]\\
&\times\left[1+\frac{i}{2\epsilon}\bm{\alpha}\cdot\partial_{\bm{\rho}}+\frac{\gamma_+}{4\epsilon}\bm{\gamma}\cdot\bm{\mathcal{A}}(T)\right]v_{p,\lambda}F(-iZ\alpha,1,i(pR_-+\bm{p}\cdot\bm{R}_-)),\quad T<0
\end{split}\\
\label{psi_mm_C}
\begin{split}
V_{C;p,\lambda}^{(\text{out})}(x)&=C_-^*\exp\left[i(pX_-)+\frac{i}{2\epsilon}\int_0^Td\tau\bm{\mathcal{A}}^2(\tau)\right]\\
&\times\left[1+\frac{i}{2\epsilon}\bm{\alpha}\cdot\partial_{\bm{\rho}}+\frac{\gamma_+}{4\epsilon}\bm{\gamma}\cdot\bm{\mathcal{A}}(T)\right]v_{p,\lambda}F(iZ\alpha,1,-i(pR_--\bm{p}\cdot\bm{R}_-)),\quad T>0
\end{split}
\end{align}
where $C_{\pm}=\exp(\pm Z\alpha/2)\,\Gamma(1\mp iZ\alpha)$, with $\Gamma(z)$ being the Gamma function, where $X_{\pm}=(t,\bm{R}_{\pm})$, with
\begin{equation}
\bm{R}_{\pm}=\bm{r}\pm\frac{1}{\epsilon}\int_0^Td\tau\bm{\mathcal{A}}(\tau),
\end{equation}
and where $F(a,b,z)$ is the confluent hypergeometric function. The states in Eqs. (\ref{psi_pp_C})-(\ref{psi_mm_C}) are the generalization of the Furry-Sommerfeld-Maue states to the case where also a plane-wave is present counterpropagating with respect to the electron \cite{Landau_b_4_1982}. We note that if one takes the two limits $|\bm{\rho}|\to\infty$ and $|T|\to\infty$ such that $|\bm{\rho}|/|T|=\text{const.}$ in the hypergeometric functions, the terms involving the laser field become negligible. This implies that the presence of the laser field does not affect the Rutherford cross section.

\section{Conclusions}
In the present paper, we have first presented a detailed derivation of the scalar and spinor Green's functions already obtained in Ref. \cite{DPM2012} in the presence of combined atomic and plane-wave fields, as well as of the correction to the total cross section of high-energy electron-positron Bethe-Heitler photoproduction in the presence of a strong laser field. In addition, we have derived for the first time the expression of the corresponding wave functions, which include exactly both the atomic and the plane-wave fields. In the particular case of the Coulomb field, these wave functions are a generalization of the Furry-Sommerfeld-Maue wave functions. With respect to the Green's functions, the availability of the wave functions offers the possibility of determining also differential cross sections of QED processes occurring in the presence of such a background field configuration.

\section*{Acknowledgments}

A. D. P. gratefully acknowledges the Budker Institute of Nuclear Physics of SB RAS for warm hospitality and partial financial support during his visit. The work  has been supported in part  by the Ministry of Education and Science of the Russian Federation and the RFBR grant no. 14-02-00016.

\appendix*
\section{}
Here, we present a derivation of the relation
\begin{equation}
\label{p_perp^2}
\mbox{e}^{-i\beta \bm p_\perp^2} g(\bm\rho)=\int\frac{d\bm{q}}{i\pi}\mbox{e}^{iq^2}g(\bm\rho+2\sqrt{\beta}\bm q)
\end{equation}
for $\beta>0$ and valid for an arbitrary functions $g(\bm{\rho})$, where $\bm{q}=(q_x,q_y)$. By writing the function $g(\bm{\rho})$ in terms of the Fourier transform $\tilde{g}(\bm{q})$ as
\begin{equation}
g(\bm{\rho})=\int d\bm{q}\,\mbox{e}^{i\bm{q}\cdot\bm{\rho}}\tilde{g}(\bm{q}),
\end{equation}
we have
\begin{equation}
\begin{split}
\mbox{e}^{-i\beta \bm p_\perp^2} g(\bm\rho)&=\mbox{e}^{-i\beta \bm p_\perp^2}\int d\bm{q}\,\mbox{e}^{i\bm{q}\cdot\bm{\rho}}\tilde{g}(\bm{q})=\int d\bm{q}\,\mbox{e}^{-i\beta q^2+i\bm{q}\cdot\bm{\rho}}\tilde{g}(\bm{q})\\
&=\int \frac{d\bm{\varrho}}{(2\pi)^2}\,g(\bm{\varrho})\int d\bm{q}\,\mbox{e}^{-i\beta q^2+i\bm{q}\cdot(\bm{\rho}-\bm{\varrho})}
=-\frac{i}{\beta}\int \frac{d\bm{\varrho}}{4\pi}\,g(\bm{\varrho})\mbox{e}^{i(\bm{\varrho}-\bm{\rho})^2/4\beta}.
\end{split}
\end{equation}
Finally, by passing from the variable $\bm{\varrho}$ to the variable $\bm{q}=(\bm{\varrho}-\bm{\rho})/2\sqrt{\beta}$, Eq. (\ref{p_perp^2}) is obtained.


\begin{thebibliography}{99}
\bibitem{Hanneke_2008}  D. Hanneke, S. Fogwell, and G. Gabrielse, Phys. Rev. Lett. \textbf{100}, 120801 (2008).


\bibitem{Sauter_1931} F. Sauter, Z. Phys. \textbf{69}, 742 (1931).
\bibitem{Heisenberg_1936} W. Heisenberg and H. Euler, Z. Phys. \textbf{98}, 714 (1936).
\bibitem{Schwinger_1951} J. Schwinger, Phys. Rev. \textbf{82}, 664 (1951).

\bibitem{Landau_b_4_1982} V. B. Berestetskii, E. M. Lifshitz, and L. P. Pitaevskii, \textit{Quantum Electrodynamics}
(Elsevier, Oxford, 1982).

\bibitem{Bethe_1934} H.~A. Bethe and W. Heitler, Proc. Roy. Soc. A \textbf{146},83 (1934).
\bibitem{Milstein_1994} A.~I. Milstein and M. Schumacher, Phys. Rep. \textbf{243}, 183 (1994).
\bibitem{Lee_2003} R.~N. Lee et~al., Phys. Rep. \textbf{373}, 213 (2003).
\bibitem{Baur_2007} G. Baur et~al., Phys. Rep. \textbf{453}, 1 (2007).
\bibitem{Baltz_2008} A.~J. Baltz et~al.,  Phys. Rep. \textbf{458}, 1 (2008).

\bibitem{Sturm_2011} S. Sturm \textit{et al.}, Phys. Rev. Lett. \textbf{107}, 023002 (2011).


\bibitem{Delbexp} Sh.Zh. Akhmadaliev, et al., Phys. Rev. C {\bf 58}, 2844 (1998).

\bibitem{splitexp} Sh. Zh. Akhmadaliev, et al., Phys. Rev. Lett. {\bf 89}, 061802 (2002).


\bibitem{Yanovsky_2008} V. P. Yanovsky \textit{et al.}, Opt. Express {\bf 16}, 2109 (2008).



\bibitem{Ritus_1985}  V. I. Ritus, J. Sov. Laser Res. \textbf{6}, 497 (1985).
\bibitem{Baier_b_1998} V.~N.~Baier, V.~M.~Katkov, and V.~M.~Strakhovenko,
\textit{Electromagnetic processes at high energies in oriented single crystals} (World Scientific, Singapore, 1998).
\bibitem{Di_Piazza_2012}  A.~Di Piazza, C. M\"{u}ller, K. Z. Hatsagortsyan, and C. H. Keitel, Rev. Mod. Phys. \textbf{84}, 1177 (2012).

\bibitem{PDG_2012} J. Beringer et al. (Particle Data Group), Phys. Rev. D \textbf{86}, 010001 (2012).
\bibitem{Wang_2013} X. Wang \textit{et al.}, Nature Commun. \textbf{4}, 1988 (2013).


\bibitem{Loetstedt_2008} E. L\"{o}tstedt et al. Phys. Rev. Lett. \textbf{101}, 203001 (2008); \textit{ibid.}, New J. Phys. \textbf{11}, 013054 (2009).

\bibitem{Di_Piazza_2009} A. Di Piazza et al., Phys. Rev. Lett. \textbf{103}, 170403 (2009); \textit{ibid.}, Phys. Rev. A \textbf{81}, 062122 (2010).



\bibitem{Loetstedt_2007} E. L\"{o}tstedt, U. D. Jentschura, and C. H. Keitel, Phys. Rev. Lett. \textbf{98}, 043002 (2007); S. Schnez et al. Phys. Rev. A \textbf{75}, 053412 (2007)

\bibitem{Di_Piazza_2008} A. Di Piazza and A. I. Milstein, Phys. Rev. A \textbf{77}, 042102 (2008).

\bibitem{DPM2012} A. Di Piazza and A. I. Milstein, Phys. Lett. B \textbf{717}, 224 (2012).


\bibitem{LM95A} R. N. Lee, A. I. Milstein, Phys. Lett. A {\bf 198}, 217 (1995); \textit{ibid.}, Zh. Eksp. Teor. Fiz. {\bf 107}, 1393 (1995) [JETP {\bf 80}, 777 (1995)].
\bibitem{LMS00} R. N. Lee, A. I. Milstein, V. M. Strakhovenko, Zh. Eksp. Teor. Fiz. {\bf 117}, 75 (2000) [JETP {\bf 90}, 66 (2000)].



\bibitem{Baier_1974} V. N. Baier, V. M. Katkov, and V. M. Strakhovenko, Zh. Exsp. Teor. Fiz. \textbf{67}, 453 (1974) [Sov. Phys.-JETP \textbf{40}, 225 (1975)].
\bibitem{Baier_1975} V. N. Baier, V. M. Katkov, and V. M. Strakhovenko, Zh. Exsp. Teor. Fiz. \textbf{68}, 405 (1975) [Sov. Phys.-JETP \textbf{41}, 198 (1975)].
\bibitem{Baier_1975_b} V. N. Baier, V. M. Katkov, A. I. Milstein, and V. M. Strakhovenko, Zh. Exsp. Teor. Fiz. \textbf{69}, 783 (1975) [Sov. Phys.-JETP \textbf{42}, 400 (1975)].
\bibitem{Baier_1975_c} V. N. Baier, A. I. Milstein, and V. M. Strakhovenko, Zh. Exsp. Teor. Fiz. \textbf{69}, 1893 (1975) [Sov. Phys.-JETP \textbf{42}, 961 (1976)].


\bibitem{Milstein_1982} A. I. Milstein and V. M. Strakhovenko, Phys. Lett. A \textbf{90}, 447 (1982).
\bibitem{Milstein_1983} A. I. Milstein and V. M. Strakhovenko, Phys. Lett. A \textbf{95}, 135 (1983); A. I. Milstein and V. M. Strakhovenko, Zh. Exsp. Teor. Fiz. \textbf{85}, 14 (1983) [JETP \textbf{58}, 8 (1983)].





\bibitem{LMS2004} R.~N. Lee, A.~I. Milstein, and V.~M. Strakhovenko, Phys. Rev. A {\bf 69}, 022708 (2004).



\bibitem{Footnote} Some numerical coefficients in the expressions of the functions $\Phi(\chi)$ and $\Psi(\chi)$ differ from the corresponding ones in \cite{DPM2012}, due to a numerical misprint in \cite{DPM2012}, which has been corrected here. The qualitative behavior of the functions $\Phi(\chi)$ and $\Psi(\chi)$ is unchanged as well as the plot in Fig. 1 and all the conclusions of Ref. \cite{DPM2012}.


\end{thebibliography}
\end{document}